\documentclass[natbib,twocolumn]{article}
\usepackage[a4paper, total={7in, 10in}]{geometry}
\pdfoutput=1

\usepackage{siunitx}
\usepackage{mathrsfs}
\usepackage{amsmath}
\usepackage{amssymb}
\usepackage{amsfonts}
\usepackage{gensymb}

\renewcommand{\mathbb}[1]{\mathbf{#1}}
\everymath{\displaystyle}

\def\cali#1{\cal#1\mit}

\usepackage{interval}
\usepackage[pdftex]{graphicx}
\usepackage[svgnames]{xcolor}
\graphicspath{{figures/}}

\usepackage{multirow}
\usepackage{booktabs,tabularx}
\usepackage{makecell}
\usepackage{array}

\DeclareMathOperator*{\nspec}{NSPEC}

\title{Rapid Scale Wind Profiling with Autoregressive Modeling and L-Band Doppler Radar}

\author{Baptiste Domps, Julien Marmain, and Charles-Antoine Gu\'erin
	\thanks{B. Domps and J. Marmain are with the Radar and Scientific Applications Department, Degreane Horizon, 83390 Cuers, France (e-mail: baptiste.domps@degreane-horizon.fr; julien.marmain@degreane-horizon.fr). C.-A. Gu\'erin is with the Mediterranean Institute of Oceanography (MIO), Universit\'e de Toulon, Aix-Marseille Universit\'e, CNRS, IRD, Toulon, France (e-mail: guerin@univ-tln.fr).}
}

\begin{document}

\maketitle

\begin{abstract}
Radar Wind Profilers (RWP) are well-established instruments for the probing of the atmospheric boundary layer, with the immense advantage of long-range and all-weather operation capability. One of their main limitations, however, is a relatively long integration time compared to other instruments such as lidars. In the context of L-band RWP we show that the use of Autoregressive (AR) modeling for the antenna signals combined with the Maximum Entropy Method (MEM) allows for a correct estimation of radial wind velocity profiles even with very short time samples. A systematical analysis of performance is made with the help of synthetic data. These numerical results are further confirmed by an experimental dataset acquired near the landing runways of Paris Charles de Gaulle (CDG) Airport, France, and validated using a colocated optical lidar at the Aerological station of Payerne, Switzerland. It is found that the AR-MEM approach can successfully derive reliable wind estimates using integration times as short as \SI{2.5}{\s} where the classical spectral approach can barely provide any measurement. In addition, the technique helps mitigating the impact of uncooperative flyers in the processing of the atmospheric echo, leading to a more successful application of the wind estimation algorithm.
\end{abstract}

\section{Introduction}

Radar wind profilers (RWP) measure vertical profiles of wind speed in the lower atmosphere under all kinds of meteorological conditions and are widely used in the context of atmospheric research (e.g. \cite{eps:dolman2018,amt:liu2020}), operational meteorology (e.g. \cite{cost:nash2001,conf:barrere2008,conf:dolman2014}), data assimilation \cite{at:wang2020} or instruments validation (\cite{acp:guo2021,tgrs:chen2020}). They generally operate in the VHF or UHF frequency bands in pulse-Doppler mode \cite{itu1997}, even though there have been isolated attempts to develop frequency modulated continuous wave (FMCW) systems (e.g. \cite{igarss:waldinger2017,igarss:klugmann2020}). They work in a coherent integration mode in order to record not only the strength of the echo but also its Doppler shift. Contrarily to optical measurement systems (lasers and lidars) they are not hindered or attenuated by clouds or fog nor necessitate the presence of aerosols as backscattering elements; contrarily to microwave weather radars, they are not bound to the sole backscattering of hydrometeors but also have the unique capability of capturing a ``clear-air'' return even in the far range, up to a few dozen of kilometers in L-band \cite{bk:zrnic1993}. In the classical electromagnetic theory, this is explained by a Bragg scattering mechanism from the small turbulent fluctuations of the refractive index in the air volume which follow the wind motion (e.g. \cite{bk:tatarski1961,proc:lane1969}). The radial speed of atmospheric echoes is characterized by a Doppler frequency peak which can be measured by a coherent integration of the backscattered signal \cite{bk:zrnic1993}. It is in general assumed that the mean velocity of the scattering structures in the atmosphere (obtained with the first moment of the Doppler spectrum) coincide with wind speed while its dispersion (quantified with the second moment) quantifies the turbulence. The wind vector can be recombined from the simultaneous measurement of radial speeds obtained with several oblique antennas, a technique often called Doppler Beam Swinging (DBS, \cite{rs:baelen1990}). Hence RWP can provide all-weather, long-range reliable estimates of the wind vector with a typical range resolution of \SIrange{150}{300}{\m} corresponding to the transmitted pulse duration.

Now, detecting the very weak clear-air echoes and accurately evaluating their speed dictates the use of ``long'' coherent integration time, that is a large number of emitted pulses. This is necessary for both increasing the Signal-to-Noise Ratio (SNR) and improving the estimation of the Doppler spectrum. Doppler spectra are then incoherently averaged over integration times of the order of \SI{20}{\s} per beam which are far beyond the temporal scale of many rapidly fluctuating phenomena such as planes and birds echoes, or vortex and wakes caused by flying objects. In part because of this limitation but also for practical reasons, RWP has been competed since the early 2000s with Doppler lidars which are compact and versatile instruments. These active laser-based instruments measure wind speed and direction at smaller spatio-temporal scales. Wind lidars indeed use much shorter integration times, of the order of \SI{1}{\s} per beam (see, e.g., \cite{jaot:grund2001}), together with smaller range cell resolution (typically \SI{50}{\m}). This has allowed for a number of studies of the turbulent atmosphere in the context of emerging industrial applications, mostly linked to the monitoring of turbulent wakes from planes or wind turbines (e.g. \cite{jaot:aitken2014,bams:nijhuis2018,ast:harris2002}). However, this is at the expense of a much shorter range, typically a few kilometers.

Nevertheless, the unique ability of RWP to study atmospheric turbulence in all-weather and the much longer range they achieve as compared to optical sensors make them irreplaceable if the limitation of long integration time can be overcome. The aim of this paper is to show that the reduced number of available samples for short integration time can be mitigated by the use of non-spectral signal-processing techniques, which bypass the time-frequency trade-off of the Fourier analysis. One such technique, based on an autoregressive (AR) representation of the backscattered signal associated with a maximum entropy method (MEM), has been successfully applied in the context of oceanographic radars \cite{joe:domps2021} to obtain Doppler spectra at rapid scale and capture some transient phenomena such as tsunamis-induced currents \cite{grsl:domps2021}. In the atmospheric context, the AR-MEM approach was successfully used for the time-frequency analysis of various transient phenomena at rapid time scale (see Figure 3 of \cite{igarss:domps2021}). Here, we apply the same method to RWP for wind measurement and demonstrate its potential on an experimental set of L-band data.

This article is organized as follows. Section \ref{sec:doppler} describes the mathematical framework used to process the atmospheric backscattered signal with the AR representation and the MEM for evaluating its coefficients. The performances of the method are assessed with synthetic RWP data and compared to a classical Fourier analysis (Section \ref{sec:synthe}). Section \ref{sec:wind_profiling} presents a first application of this methodology to an actual RWP dataset. Finally, the validity of the short-time estimations is evaluated against lidar measurements (Section \ref{sec:lidar}).

\section{AR Modeling of the Atmospheric Doppler Spectrum}\label{sec:doppler}

We briefly recall the backscattering mechanisms at play in the turbulent atmosphere and the few signal processing steps which are conventionally employed to extract the spectral moments from the antenna voltage time-series. The reader is referred to \cite{rs:woodman1985,ageo:barth1994,rs:carter1995} for a complete review of the processing scheme.

\subsection{Atmospheric Doppler Spectrum}\label{sec:doppler_spectrum}

The theoretical Doppler spectrum $\sigma(\omega)$ is the limiting form of the power spectral density (PSD) of the backscattered range-resolved complex time-series $s(t,r)$ for large integration time and a given detection volume (the range index $r$ is implicit and will be omitted in the following). To an amplitude factor, given by the radar equation, the Doppler spectrum writes
\begin{equation}
	\label{eq:doppler_spectrum}
	\sigma(\omega) \sim \Bigg<\bigg|\int_0^Te^{-i\omega t}s(t)W(t)\,dt\bigg|^2\Bigg>
\end{equation}
where $W(t)$ is a windowing function over the integration time $T$ and $\omega$ is the circular frequency. As it is well known, the Doppler frequency shift is proportional to the radial velocity $V_r$ of scatterers, $\omega=2\pi f=4\pi V_r/\lambda$, where $\lambda$ is the radar wavelength. Hence the Doppler spectrum can be interpreted as a power-weighted distribution of the radial velocities within the resolution volume \cite{bk:zrnic1993}, where the weighting function depends on the scatterers reflectivity and velocity. In the following we will therefore express the Doppler spectra as a function of the radial speed $v$ rather than Doppler frequency Due to the random motion of scatterers during the observation time, the shape of the weighting function is generally assumed to be Gaussian and centered about the average radial wind speed. Its width is related to the distribution of the individual scatterer velocities which is spread by shear or turbulence. It can happen that the assumption of Gaussian distribution of velocities is violated when some strong intermittent clutter echoes (in particular from flyers) is superimposed to the atmospheric return. Such parasitic echoes should be filtered or at least ignored when inferring radial wind speed from Doppler spectrum. Therefore in the absence of noise and clutter, a Gaussian-shaped atmospheric Doppler spectrum has the generic form \cite{rs:woodman1985}:
\begin{equation}
	\label{eq:gaussian_spectrum}
	S(v) = \frac{S_0}{\sigma_v\sqrt{2\pi}} \exp\left(-\frac{(v-V_r)^2}{2\sigma_v^2}\right)
\end{equation}
Under this form, the Doppler spectrum  is fully characterized by its first three moments:
\begin{equation}
	{\cali M}_n=\int_{-\infty}^{+\infty} v^nS(v)\,dv,\ n=0,1,2
\end{equation}
The zeroth-moment $S_0={\cali M}_0$ is the weather signal power, the normalized first-moment $V_r={\cali M}_1/{\cali M}_0$ is the mean radial velocity and the normalized second-moment $\sigma_v^2={\cali M}_2/{\cali M}_0$ is the radial velocity variance.


Complex voltage time-series are coherently recorded at the output of a quadrature demodulator and sorted into range cells. For each gate, around 50 to 100 samples (that is 50 to 100 emitted pulses) are coherently summed which is equivalent to a time-domain averaging. After digital decimation the sampling frequency is of the order of 150 to \SI{200}{\hertz} giving a typical Nyquist interval of the order of $\pm\SI{10}{\m\per\s}$. Note that the sampling strategy can be adapted by the radar operators depending on the atmospheric conditions. The calculation of the Doppler spectrum (\ref{eq:doppler_spectrum}) is usually performed with a Fast Fourier Transform (FFT) over short subseries (typically, $N=128$ time steps):
\begin{equation}
	\label{eq:psd_fft}
	P(\omega_k) = (\Delta t)^2\left|\sum_{n=0}^{N-1}e^{-i\omega_kn}s(n\Delta t)W(n\Delta t)\right|^2
\end{equation}
where $\omega_k=k/T$ for $k=0,\dotsc,N-1$ is the discrete set of dual frequencies and $\Delta t$ is the sampling period after coherent integration and digital decimation.

The resulting sample spectra are further incoherently summed over a certain number $\nspec$ of half-overlapping sequences to obtain the ensemble average (\ref{eq:doppler_spectrum}). Incoherent averaging is performed using the Statistical Averaging Method (SAM) of Merritt \cite{jaot:merritt1995}. In the SAM approach, it is assumed that the atmospheric signal has Gaussian statistics whereas intermittent echoes (birds, planes, uncooperative flyers, etc.) are larger, non-Gaussian signals. Temporal variations of each range-Doppler bin (or, equivalently, range-radial velocity bin) should therefore comply to an exponential distribution test as proposed by Hildebrand and Sekhon (HS) \cite{jam:hildebrand1974}. Hence the SAM makes only use of the frequency bin passing the exponential energy distribution test of HS, every other being considered as non-atmospheric echoes and discarded from the average. The main strength of the SAM is that it needs no assumption on the nature of the intermittent echoes. However, as it is based on a statistical criterion, the SAM assumes that the atmosphere can be observed free of parasitic echoes at each range gate and frequency bin for at least one sample Doppler spectrum.


Once the ``clean'' Doppler spectra are obtained, a Multipeaks Picking Procedure (MPP) is used (see e.g. \cite{jaot:cornman1998,jaot:morse2002,amt:williams2018}) to identify at most three peaks for each range gate as the averaged spectra may still contain some ground-clutter and flyers echoes in addition to the atmospheric signal. Selection of one single peak would therefore lead to wrong estimates as non-atmospheric echoes may remain after the SAM processing. The first three spectral moments of the identified  peaks are computed and used to remove those having specific non-atmospheric signatures. Finally, some empirical criteria on the spectral moments are used along with the constraint of velocity continuity with respect to range to select for the most likely atmospheric peak. In the following, we will refer to this standard processing scheme as the ``FFT method''.
    
\begin{figure}[h]
	\centering
	\includegraphics[width=\linewidth]{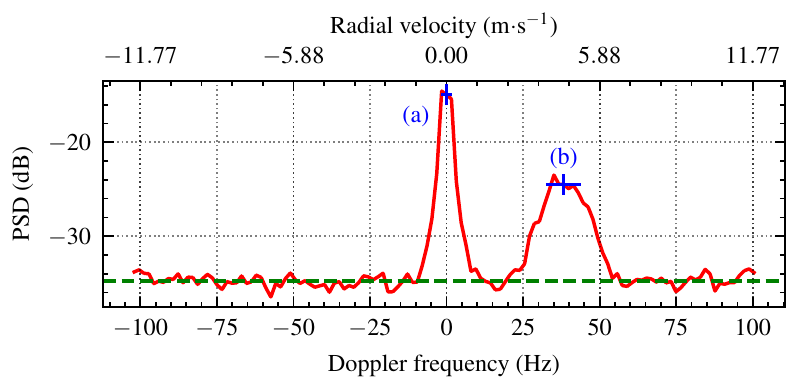}
	\caption{Example of typical Doppler spectrum (\si{\deci\bel}) obtained using an L-band RWP and standard processing on range gate 14 of ``low mode'' (\SI{1050}{\meter} range). Blue crosses are centered on two peaks identified by the multipeaks picking procedure (MPP): (a)~Ground clutter; (b)~Atmospheric echo. For each peak, length of the horizontal bars correspond to the estimated spectral width. Dashed line marks the noise level obtained by the Hildebrand and Sekhon (HS) method.}
	\label{fig:typical_doppler}
\end{figure}

A typical RWP Doppler spectrum for a resolution volume in fair weather is shown in Figure \ref{fig:typical_doppler}. It is obtained from an incoherent summation of $\nspec=61$ instantaneous spectra of $128$ samples each. The sampling frequency is \SI{204}{\hertz}, leading to an unambiguous (or Nyquist) velocity of $N_v=\SI{12}{\m\per\s}$. Two echoes, labeled (a) and (b), can be observed in the spectrum. The zero-velocity peak corresponds to reflections from the ground or ``fixed-echoes'', while the broader peak is a Gaussian-shaped atmospheric echo. Here, fixed-echoes are of greater energy than the atmospheric return, a situation commonly encountered in the lowest range gates. The mean radial velocity is estimated to $\hat{V}_r=\SI{4.50}{\m\per\s}$ and the spectrum width to $\hat{\sigma}_v=\SI{1.36}{\m\per\s}$. Note that the spectrum width is often normalized by twice the Nyquist velocity, so that $\hat{\sigma}_{vn}=\hat{\sigma}_v/2N_v=0.057$.

The temporal resolution of the inverted wind velocity is constrained by the minimal integration time $T=N\Delta t$ of each individual Doppler spectra necessary to ensure a satisfactory frequency (hence radial velocity) resolution and SNR and a sufficient number $\nspec$ of incoherent spectra to obtain a meaningful ensemble average. With the typical experimental values $T=128\times \SI{5}{\milli\s}=\SI{640}{\milli\s}$ for the coherent integration time and $50$ for the number of incoherent spectra, this leads to a temporal resolution of about 16 seconds for the wind velocity, an acquisition rate that might be too coarse for the observations of some transitory phenomena. Due to the time-frequency dilemma, the integration time cannot be further reduced in the frame of the FFT method. This limits the accuracy of the velocity estimation and calls for an alternative, non-spectral method.

\subsection{AR-MEM Representation of Time Series and PSD}

In the AR approach, the instantaneous received complex signal at the sample rate $\Delta t$ is modeled as a linear combination of its past values and a white noise $\varepsilon$
\begin{equation}
	\label{eq:mod_ar}
	s(n\Delta t) = -\sum_{k=1}^pa[k]s\left(s(n-k)\Delta t\right)+\varepsilon[n]
\end{equation}
The number $p$ of values in the past which are involved in the summation is termed the AR order and the $\{a[k]\}_{k=1}^p$ are the AR coefficients. After determination of the latter, the signal PSD is given by \cite{bk:stoica2005}:
\begin{equation}
	\label{eq:psd_ar}
	P(\omega) = P_\varepsilon\left|1+\sum_{k=1}^pa[k]e^{-i\omega k\Delta t}\right|^{-2}
\end{equation}
where $P_\epsilon$ is the white noise PSD. It must be underlined that this derivation of the PSD presents important differences with respect to the expression (\ref{eq:psd_fft}) obtained with a classical FFT. In the AR approach, the signal is not assumed to be periodic and the spectral estimation does not suffer from windowing nor truncation. Furthermore, the PSD (\ref{eq:psd_ar}) can be evaluated at arbitrary frequency $\omega$ and is not bound to a discrete set of dual frequencies as imposed by the FFT. As pointed out by \cite{med:kaluzynski1989}, this is particularly suitable for short time series, when the FFT would neither ensure sufficient frequency resolution nor statistical convergence.

A key point in estimating the AR spectra is the choice of the optimal order $p$, a problem that has been extensively discussed in the literature (e.g. \cite{joe:paulson1987,tsp:waele2003}). Small values of $p$ with respect to the number of samples $N$ generally leads to a poor frequency resolution while large values can induce spurious spectral peaks. It has therefore been suggested by \cite{bk:ulrych1979} to select $N/3-1 \le p \le N/2-1$. In the specific context of radar applications, the authors proposed to retain $p=N/2$ \cite{joe:domps2021} on the basis of empirical tests.

According to (\ref{eq:psd_ar}), estimating the AR PSD is equivalent to finding the AR coefficients. Multiple methods have been proposed, namely the Yule-Walker equations, the Levinson-Durbin algorithm, or the Burg method (we refer to \cite{bk:marple2019} for a review). This last method is known to give PSD estimates with high resolution, especially for short records \cite{jpe:saito1978,bk:claerbout1985}, a property which motivated our choice in the present context. In the Burg method \cite{phd:burg1975} the AR coefficients are obtained by minimizing the square sum of the forward \textit{and} backward prediction errors, considering the AR model (\ref{eq:mod_ar}) as a linear prediction filter. Multiple implementations of the Burg's method are nowadays available in most digital signal processing programming languages. The Burg's method can further be interpreted in terms of signal entropy as it maximizes the randomness of the unknown time series. In agreement with the radar literature \cite{bk:heidbreder1991}, we will hereinafter refer to the Burg's method as the Maximum Entropy Method (MEM) and its combination with an AR process as the AR-MEM. The standard processing scheme described in Section \ref{sec:doppler_spectrum} with the replacement of the FFT with the AR-MEM for the computation of the Doppler spectrum will be referred to in the following as the ``AR-MEM method''.

\section{Evaluation With Synthetic Data}\label{sec:synthe}

To assess the performances of the AR-MEM in the context of rapid scale wind profiling, we went through a series of tests using synthetic RWP-like time-series with a known radial wind velocity. We used the method described in  \cite{jam:zrnic1975} for constructing synthetic data with realistic statistical properties. We could therefore generate ``weatherlike'' voltage complex time series with prescribed radial wind speed $V_r$, spectral width $\sigma_{vn}$ and SNR. To do this we first construct the signal PSD as:
\begin{equation}
	\label{eq:typical_spectrum}
	P(\omega)=-\big(\sigma(\omega)+C(\omega)+\mathcal{N}\big)X(\omega)
\end{equation}
where $\sigma$ is a centered Gaussian-shaped atmospheric PSD (\ref{eq:gaussian_spectrum}), $C$ is a narrow Gaussian-shaped PSD accounting for ground-clutter echo, $\mathcal{N}$ is the uniform normalized PSD of white noise and $X$ is an exponentially distributed random variable. The complex spectral components lead to the complex voltage time series through an inverse Fourier transform:
\begin{equation}
	\label{eq:radar_sim}
	s(t) = \sum_j \sqrt{P(\omega_j)} e^{i(\omega_j t+ \varphi_j)} e^{i\Phi_D(t)}
\end{equation}
where $\omega_j=2\pi j/T$ and $\varphi_j$ are uniform independent random phases. By construction, the amplitudes $\sqrt{P}$ are Rayleigh distributed and the individual frequency components are complex Gaussian variables. The deterministic varying phase $\Phi_D(t)$ accounts for the phase shift induced by the velocity of the inhomogeneities carried by the wind $\Phi_D(t)=4\pi/\lambda V_rt=\omega_Dt$.

\begin{figure}[t!]
	\centering
	\includegraphics[width=\linewidth]{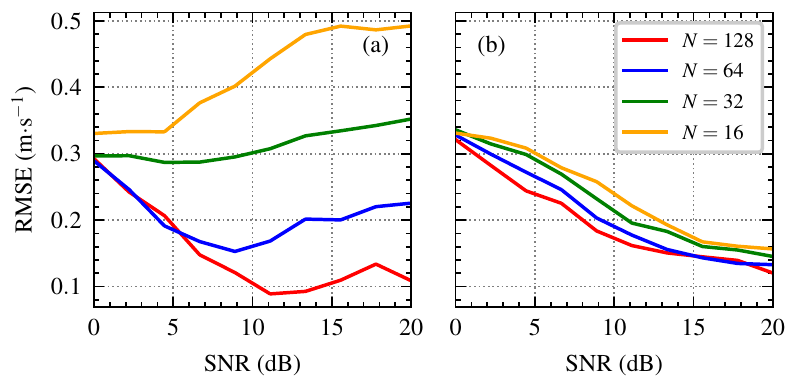}
	\caption{Influence of the SNR (\si{\deci\bel}) on the RMSE (\si{\m\per\s}) in estimating the first-order moment (radial wind speed $U_r$) from synthetic RWP time-series with $U_r=\SI{4}{\m\per\s}$ and using: (a)~FFT; (b)~AR-MEM. The radar signal is modeled by a random stationary process with prescribed Doppler spectrum (\ref{eq:radar_sim}) and a representative value of spectral width ($\sigma_{vn}=0.313$). For each SNR value, 100 time series are generated and split into $60$ overlapping sub-series of $N$ time steps.}
	\label{fig:snr_theorique}
	\vspace{2mm}
	\includegraphics[width=\linewidth]{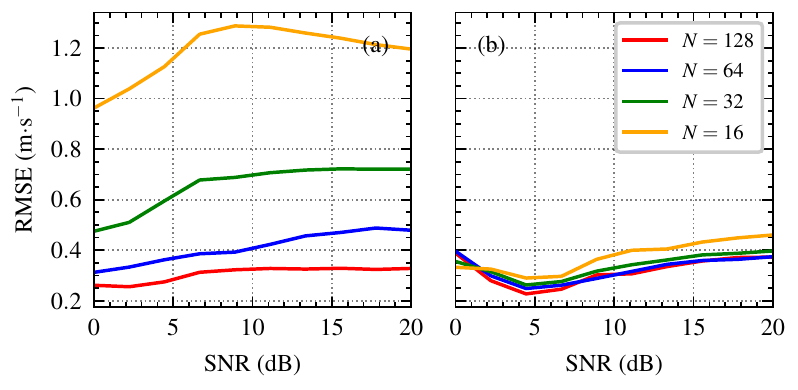}
	\caption{Same as Figure \ref{fig:snr_theorique} but for the RMSE (\si{\m\per\s}) in estimating the second-order moment (spectral width $\sigma_v$).}
	\label{fig:largeur_theorique}
\end{figure}

We performed a series of numerical tests according to the model (\ref{eq:typical_spectrum}), (\ref{eq:radar_sim}) and (\ref{eq:gaussian_spectrum}) for some typical values of the radial wind speed $V_r$ and spectral width $\sigma_v$ and for a wide range of samples sizes $N$ and SNR. The instantaneous PSD was recalculated using either the AR-MEM or the classical FFT approach by processing half overlapping series of variable length. The sample size was increased from $N=16$ to $N=128$ corresponding to an integration time from $T=80$~ms to $T=640$~ms. While the number of available frequency bins with the FFT is equal to the sample length, the AR-MEM makes it possible to evaluate the PSD using (\ref{eq:psd_ar}) over an arbitrary large number of frequency bins (here 128). For each test case, a total of 100 random time series were generated and the root-mean-square error (RMSE) was calculated between estimated and simulated radial wind speed. Figure \ref{fig:snr_theorique} shows the influence of the SNR on the accuracy of the first moment for a radial wind speed of \SI{4}{\meter\per\s} and a typical spectral width ($\sigma_{vn}=0.0313$ \cite{jaot:may1989}) when using FFT (Figure \ref{fig:snr_theorique}(a)) and AR-MEM (Figure \ref{fig:snr_theorique}(b)). As expected, the RMSE is smaller for ``long'' than ``short'' integration times in both FFT and AR-MEM. Furthermore, in the case of ``short'' integration times ($N=16$), the RMSE in AR-MEM is twice lower than in FFT (\SI{0.56}{\meter\per\s} versus \SI{1.01}{\m\per\s}, respectively). Note that the quality of the estimation is robust to noise, as seen from the quasi-constant value of the RMSE as the SNR is increased. This is due to the barycentric method used by the MPP to identify the precise peak location within the Doppler spectrum. In the same manner we also assessed the robustness of the second-order moment (Figure \ref{fig:largeur_theorique}). For ``long'' time series ($N=128$) FFT and AR-MEM provide accurate estimates, with an RMSE close to \SI{25}{\cm\per\s}. However, the RMSE of the estimation obtained with FFT increases significantly when the sample size is decreased (up to \SI{1.5}{\m\per\s} for $N=16$). As previously stated the frequency resolution of the FFT is insufficient for such short samples, causing the atmospheric echo to spread over multiple frequency bins and mixing with the ground-clutter echo. In AR-MEM, the frequency resolution remains constant regardless of the integration time and the atmospheric echo is adequately described in the Doppler spectrum.

Finally we assessed the computational time needed to produce AR-MEM Doppler spectra with respect to the conventional FFT method. The latter is known to require $O(N\log N)$ operations for a single sample of $N$ time steps. In AR-MEM, the authors estimated that the computational time with an optimal order $p=N/2$ is of the order of $O(N^2)$ \cite{joe:domps2021}. This should not prevent from using the AR-MEM as the average computational time per beam for 60 range gates (that is, the average computational time requested for computing Doppler spectra as Figure \ref{fig:compare_psd}(c)) was found to be \SI{21.40}{\s} on a standard desktop computer, while being only of \SI{2.18}{\s} in FFT. In practice, this computational time can be further decreased by diminishing the order ($N/3$ instead of $N/2$) without significantly deteriorating the spectral estimation.

\section{Experimental assessment with a L-band RWP}\label{sec:wind_profiling}

The potential of the AR-MEM to estimate wind velocities at a rapid time scale in experimental conditions was evaluated on the basis of the measurements provided by the PCL1300 UHF RWP manufactured by company Degreane Horizon, operating at \SI{1.274}{\giga\hertz}. The data set under consideration was acquired on September 24, 2012, near the landing runways of Paris Charles de Gaulle Airport, Roissy-en-France, France (\SI{49}{\degree N}, \SI{2.75}{\degree E}, \SI{107}{\m} above sea level). The weather was mostly cloudy with an average relative humidity of \SI{76}{\%}. Four radar beams were pointed $\theta=\SI{17}{\degree}$ from the zenith to the north, east, west and south and a fifth beam pointed to the zenith, using the classical DBS technique. Uninterrupted time series of 3264~samples (\SI{17.5}{\s}) were collected for each beam and range gate. In order to achieve a trade-off between the range resolution and the available range, which are impacted in contradictory ways by the pulse duration, two pulsing modes were used alternatively. The first sweep cycle was realized on the 5 antennas in ``high mode'', that is with a pulse duration of $\tau=\SI{2.5}{\micro\s}$; a second sweep cycle was run in ``low mode'' corresponding to a shorter pulse duration of $\tau=\SI{1}{\micro\s}$. The high mode provides measurements to a maximal theoretical range of \SI{8.5}{\kilo\m} with a range resolution of \SI{375}{\m} while the low mode give access to a finer range resolution (\SI{150}{\m}) over a shorter range interval (up to \SI{4.5}{\kilo\m}). Hereinafter, the ``reference'' processing scenario consists in using the FFT method over ``long'' half overlapping samples ($N=128$) and applying the SAM over the resulting $\nspec=61$ individual PSD. This corresponds to a total integration time of \SI{17.5}{\s} and provides a frequency resolution $\Delta f = \SI{1.61}{\hertz}$ allowing for a velocity resolution $\Delta V_r = \SI{0.19}{\m\per\s}$, assuming stationary wind over this duration.

\begin{figure}[t]
	\includegraphics[width=\linewidth]{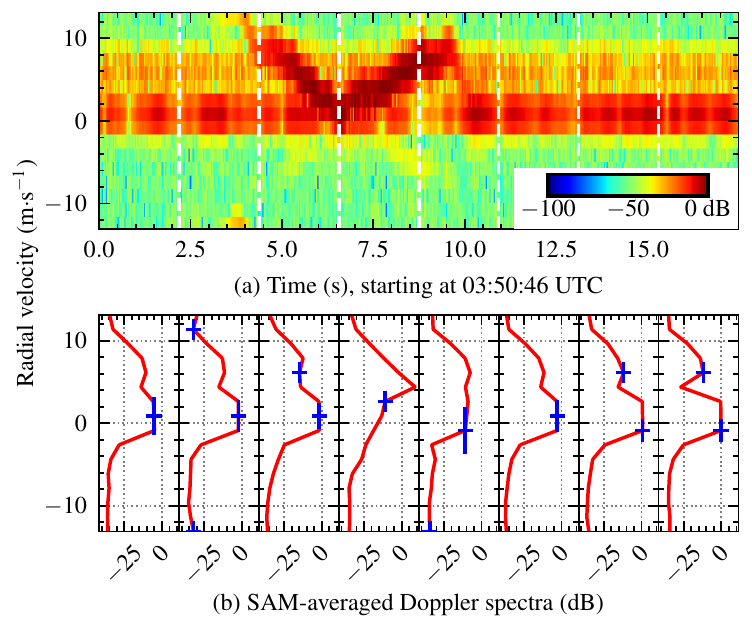}
	\caption{Data recorded by the Paris CDG RWP on September 24, 2012, on the range gate 3 of ``low mode '' (\SI{225}{\m} range) and processed using the FFT method from ``short'' time-series ($N=16$ time steps). (a) Time-frequency spectrogram (\si{\deci\bel}, colorscale); (b) SAM-averaged Doppler spectra (\si{\deci\bel}) obtained using the Merritt algorithm applied over $\nspec=61$ individual spectra.}
	\label{fig:tf_fft}
\end{figure}

\begin{figure}[h]
	\includegraphics[width=\linewidth]{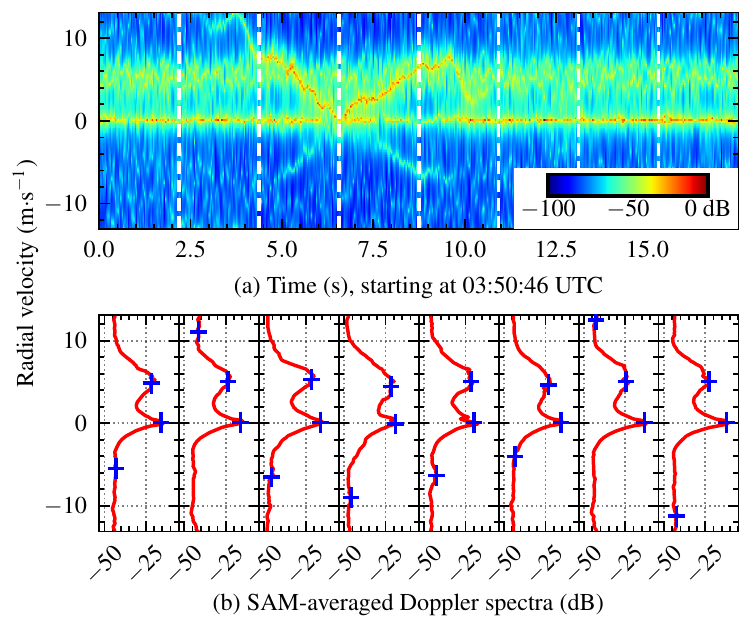}
	\caption{Same as Figure \ref{fig:tf_fft}, but using the AR-MEM. IQ-imbalance artifacts can be seen in the negative radial velocities of the panels three to five of the time-frequency spectrogram.}
	\label{fig:tf_ar}
\end{figure}

We first assessed the performances of the AR-MEM in computing the Doppler spectrum from ``short'' time-series. In this fast-time scenario the voltage time-series were split into 487 overlapping sub-series of $N=16$ time steps from which the individual PSD were computed using the FFT and the AR-MEM. As an example the temporal variations of the instantaneous PSD are shown in the time-Doppler plane in Figs. \ref{fig:tf_fft}(a) and \ref{fig:tf_ar}(a) using \SI{17.5}{\s} of data. The FFT spectrogram yields a coarse frequency resolution $\Delta f = \SI{13.6}{\hertz}$ (or velocity resolution $\Delta V_r = \SI{1.6}{\m\per\s}$), while the short-time AR-MEM computed with optimal order ($p=8$) on a finer frequency grid (128 bins) exhibits a frequency resolution $\Delta f = \SI{1.76}{\hertz}$ (that is $\Delta V_r = \SI{0.21}{\m\per\s}$). In the former case (Figure \ref{fig:tf_fft}(a)), the wind echo in the positive radial velocities is mixed with the ground-clutter. A non-cooperative flyer, likely a bird, produces a strong echo between the 5th and 9th second on the time axis and cannot be distinguished from the atmospheric return. However, the AR-MEM time-Doppler spectrogram (Figure \ref{fig:tf_ar}(a)) preserves an accurate description of the atmospheric echo, even in presence of the flyer.

\begin{figure*}[h]
	\centering
	\includegraphics[width=.9\linewidth]{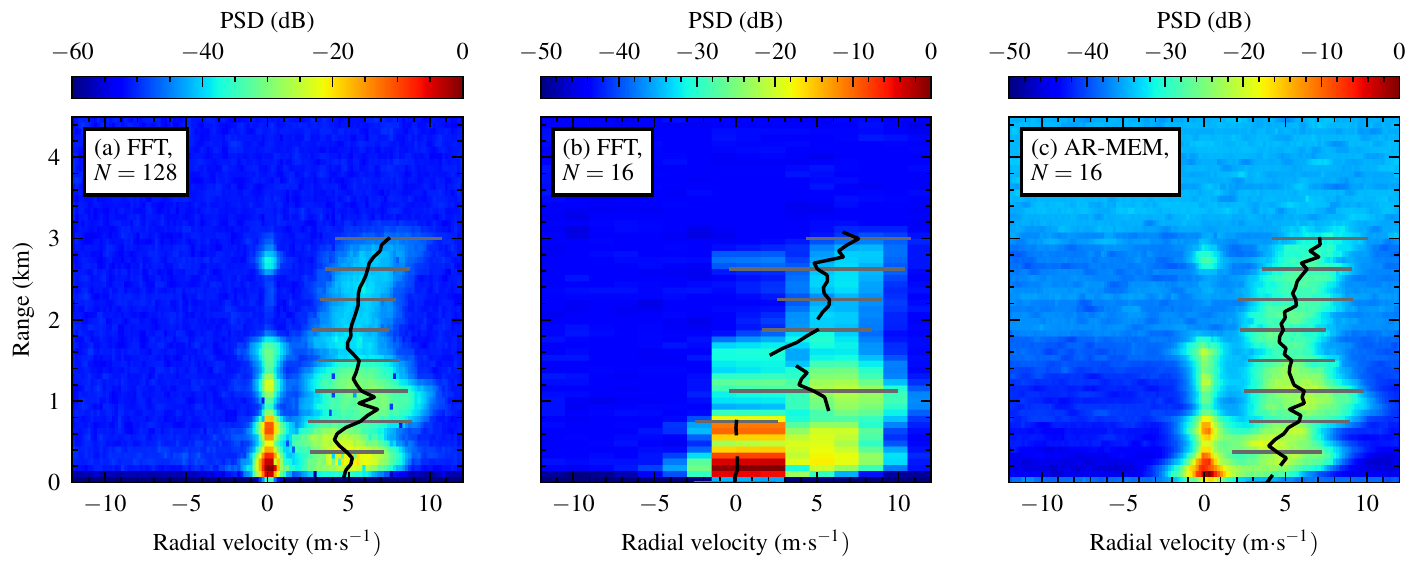}
	\caption{Normalized Range-Doppler power spectra (\si{\deci\bel}; colorscale), inferred wind profiles (\textbf{----}, \si{\m\per\s}) and first-order spectral width ({\color{DimGray}\textbf{----}}, \si{\m\per\s}). The acquisition was performed on September 24, 2012, at 19:03:27~UTC along the southern beam of the Paris CDG RWP operating in low mode, and processed using: (a)~FFT and long sub-series ($N=128$); (b)~FFT and short sub-series ($N=16$); and (c)~AR-MEM and short subseries ($N=16$). Incoherent averaging is set to a maximum of $60$ subseries. Solid lines represent the first-order spectral moments obtained using MPP and horizontal bars mark the second-order spectral moment.}
	\label{fig:compare_psd}
\end{figure*}

\begin{figure*}[h]
	\centering
	\includegraphics[width=.8\linewidth]{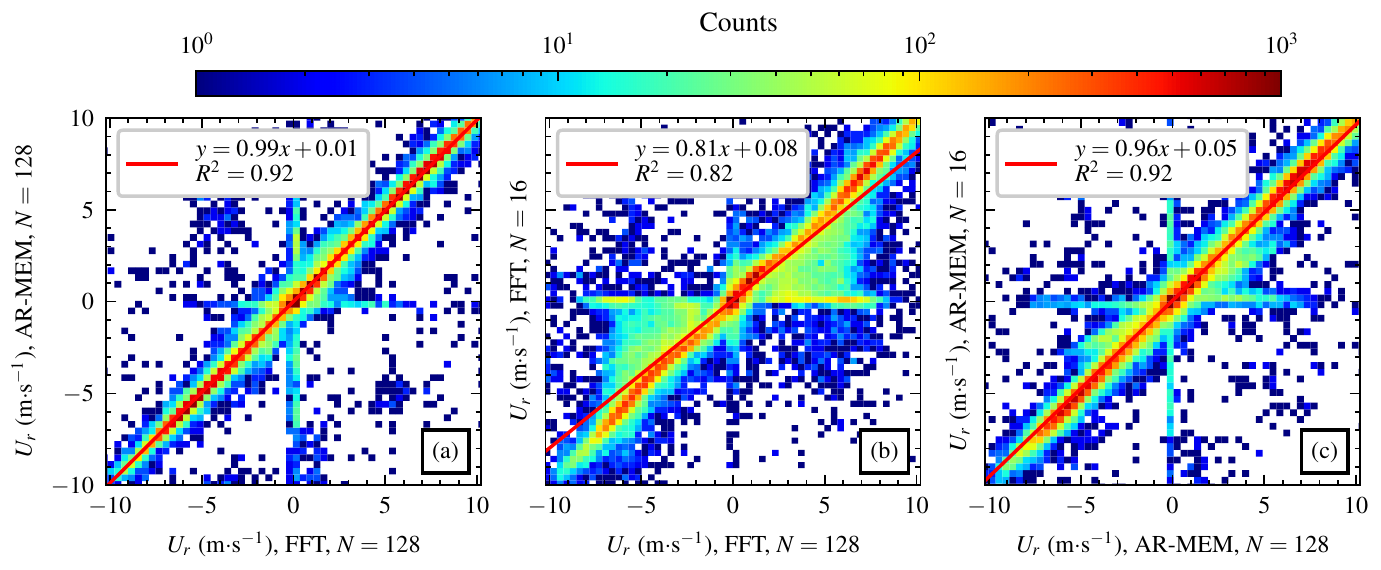}
	\caption{Scatter plots (colorscale, counts) of estimated first-order moments (or radial velocity $U_r$, \si{\m\per\s}) along every range gate of the southern beam of the Paris CDG RWP on September 24, 2012, operating in low mode and during a 24-h interval. (a)~FFT and long subseries ($N=128$) versus AR-MEM and long subseries ($N=128$). (b)~FFT and long subseries ($N=128$) versus FFT and short subseries ($N=16$). (c)~AR-MEM and long subseries ($N=128$) versus AR-MEM and short subseries ($N=16$). The RMSE (with respect to the reference long-time estimates) using short-time FFT was found to be \SI{4.83}{\m\per\s}, while being only of \SI{2.10}{\m\per\s} in short-time AR-MEM.}
	\label{fig:compare_estim}
\end{figure*}

To confirm these observations we further used the SAM from 8 groups of $\nspec=61$ individual PSD (delimited by dashed lines on Figs. \ref{fig:tf_fft}(a) and \ref{fig:tf_ar}(a)). The resulting Doppler spectra are shown in Figs. \ref{fig:tf_ar}(a) and \ref{fig:tf_ar}(b) for the FFT and AR-MEM methods, respectively. In the fast-time scenario using FFT (Fig. \ref{fig:tf_ar}(b)), the atmospheric echo is mixed with the ground-clutter for the near-zero velocities. The MPP therefore fails to retrieve the wind peak as the resolution of the Doppler spectrum is clearly insufficient. However the AR-MEM Doppler spectra (Fig. \ref{fig:tf_fft}(b)) is able to identify unambiguously two distinct peaks corresponding to the ground-clutter and the atmospheric return.



We now assess the performances of AR-MEM with respect to the FFT method by comparing the corresponding range-Doppler profiles. Fig. \ref{fig:compare_psd} depicts such profiles along the southern beam for the high mode data using either method. Again, the PSD was computed using $\nspec=61$ half overlapping time-series followed by the SAM as previously described. The MPP is applied on each range gate to identify 3 peaks and a consensus algorithm is used to retain the most probable wind peak. 
Both the aforementioned reference and fast-time scenarios are tested with the FFT and AR-MEM methods. In the former scenario (Fig. \ref{fig:compare_psd}(a)), a typical wind echo is clearly identifiable up to \SI{3}{\kilo\meter} with a radial velocity close to \SI{5}{\m\per\s} and mean spectral width of \SI{2.5}{\m\per\s}. In the fast-time scenario using FFT (Fig. \ref{fig:compare_psd}(b)), the atmospheric echo is mixed with the ground-clutter in the lowest range gates. The spectral moments could not be properly retrieved as the resolution of the Doppler spectrum is very insufficient, leading to obviously erroneous wind estimates. However, the short-time AR-MEM (Fig. \ref{fig:compare_psd}(c)) computed with optimal order ($p=8$) on a finer frequency grid (128 bins) leads to a wind profile which is fully consistent with the reference scenario.

To confirm this observation, a systematic comparison of the two processing methods using long and short integration time was run using \SI{24}{\hour} of the high-mode data acquired along the 5 beams. We performed consistency tests between the velocity estimates obtained with long and short integration times and obtained using the different methods. The scatterplots in Fig. \ref{fig:compare_estim} show that the FFT and AR-MEM method are perfectly consistent for long integration times but quite different for short integration time, where only the AR-MEM is capable of reproducing the performances of the long time counterpart.

\begin{figure}[h]
	\centering
	\includegraphics[width=\linewidth]{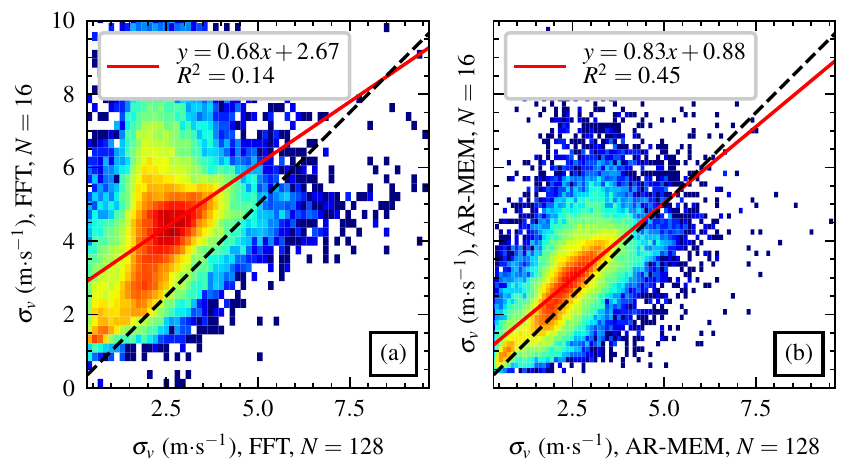}
	\caption{Scatter plots (same colorscale as Fig. \ref{fig:compare_estim}) of estimated second-order moments (in terms of velocity spread $\sigma_v$, \si{\cm\per\s}) from the same dataset as Figure \ref{fig:compare_estim} and processed using: (a)~FFT and long subseries ($N=128$) versus AR-MEM and long subseries ($N=128$). (b)~FFT and long subseries versus FFT and short subseries ($N=16$). (c)~AR-MEM and long subseries versus AR and short subseries. In any case, incoherent averaging is set to a maximum of $60$ subseries.}
	\label{fig:compare_largeur}
\end{figure}

In a similar way we investigated the accuracy of the second-order moment $\sigma_v$. As shown in Fig. \ref{fig:compare_largeur}(a), the spectral width parameter is significantly overestimated using the short-time FFT when compared to the reference long-time FFT estimates, with a RMSE of \SI{6}{\m\per\s}. Again, the AR-MEM  estimates benefit from a finer frequency resolution of the PSD and have a much smaller RMSE (\SI{0.88}{\m\per\s}).

\section{Validation with colocated Lidar measurements}\label{sec:lidar}

The validity of the short-time AR-MEM to infer radial wind velocities from RWP experimental measurements has first been assessed by self-consistency tests with long-time estimates (Figures \ref{fig:compare_estim} and \ref{fig:compare_largeur}). We now provide a further validation of the method by comparison with a colocated wind Doppler lidar. The comparison of radar and lidar profiler is always challenging due to the very different spatio-temporal scales  entering in play,. However, a good agreement between the two types of instruments has already been reported in the literature. For example using the NOAA high-resolution Doppler lidar (HRDL) and a \SI{915}{\mega\hertz} boundary layer RWP, \cite{jaot:cohn2002}, an excellent correlation between lidar and RWP observations ($R^2=0.99$) and a small mean absolute error (\SI{0.15}{\m\per\s}) were reported.

The integration times required with lidars are by at least one order of magnitude smaller than those typically employed with RWP; this is therefore adapted to evaluate the accuracy of short-time measurements with RWP. Both instruments were operated at the Aerological station of Payerne (\SI{7}{\degree E}, \SI{47}{\degree N}, \SI{491}{\m} above sea level), \SI{45}{\kilo\meter} North-East of Lausanne, Switzerland. This station is part of a network owned by the Federal Office of Meteorology and Climatology (MeteoSwiss) intended to monitor the atmosphere above the Swiss Plateau. The RWP of Payerne has the same characteristics as the aforementioned Paris CDG RWP (see Section \ref{sec:wind_profiling}), except that its oblique beams are rotated by \SI{45}{\degree} with respect to North, hence pointing to North-East, South-East, South-West and North-West, respectively, with an angle $\theta=\SI{17}{\degree}$ to the zenith. The lidar is of type WindCube WLS200S manufactured by company Vaisala. An overview of the operating characteristics of this lidar, compared to the RWP, is shown in Table \ref{tab:radar_vs_lidar}. The measurements were performed under clear weather from 12:00 to 23:59 UTC on March 2, 2021.

\begin{table}[h]
	\centering
	\caption{Characteristics for the Degreane Horizon PCL1300 RWP and the Vaisala WindCube WLS200S Doppler lidar}
	\label{tab:radar_vs_lidar}
	\begin{tabular}{ccc}
		\toprule
		Feature & Wind profiler & Doppler lidar \\
		\midrule
		Wavelength & \SI{23}{\centi\m} & \SI{1.54}{\micro\m} \\
		Vertical resolution & \SI{150}{\m} & \SI{30}{\m} \\
		Pulse duration & \SI{1}{\micro\s} & \SI{200}{\nano\s} \\
		Typical integration time & \SI{15}{\s} & \SI{0.5}{\s} \\
		\bottomrule
	\end{tabular}
\end{table}

\begin{figure*}[t]
	\centering
	\includegraphics[width=\linewidth]{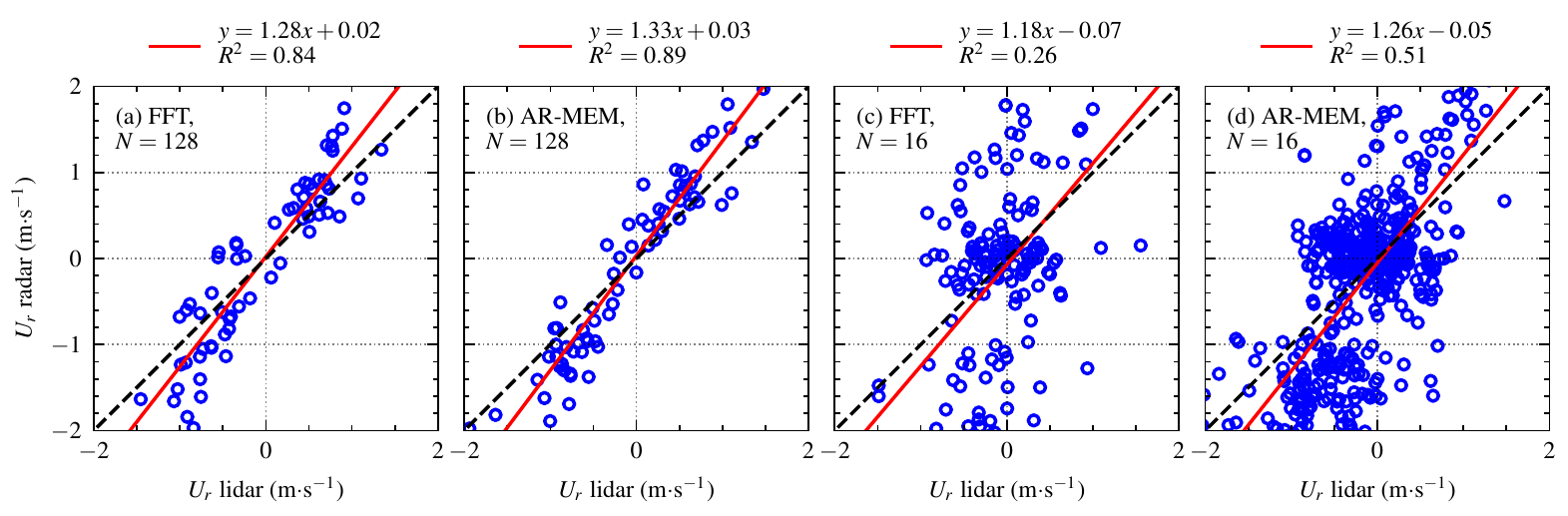}
	\caption{Scatter plots of estimated radial wind velocity during \SI{12}{\hour} on March 2, 2021, using the RWP of Payerne and a colocated lidar as reference. (a) FFT, long sample ($N=128$ time steps). (b) AR-MEM, long sample ($N=128$ time steps). (c) FFT, short sample ($N=16$ time steps). (d) AR-MEM, short sample ($N=16$ time steps). In any case, incoherent averaging is set to a total of 60 subseries.}
	\label{fig:compare_radar_lidar}
\end{figure*}

For the purpose of the comparison with the RWP, the lidar data under consideration were processed as follows. First, the zonal, meridional and vertical wind components $(u,v,w)$ were projected along the four oblique radar beams as
\begin{equation}
	\label{eq:proj_lidar}
	U_{r,\varphi} = u\sin\theta\sin\varphi + v\sin\theta\sin\varphi + w\cos\theta
\end{equation}
where $\varphi$ is the beam azimuth with respect to North. Then the projected radial velocities were interpolated on the radar range gates and the lidar measurements contained within each radar integration time were averaged. The comparisons were made using FFT and AR-MEM methods, using long ($N=128$ time steps) and short ($N=16$ time steps) time series. We chose to present only the measurements performed on two opposite radar beams, here the North-East and the South-West, but similar results were obtained using the two remaining beams.

The results of the comparisons for each different processing are shown in Figure \ref{fig:compare_radar_lidar} using the lidar measurements as reference. Additionally, the performances of the different methods when compared with the reference lidar measurements are recap in Table \ref{tab:compare}. For long time-series, the FFT and AR-MEM estimates (Figs. \ref{fig:compare_radar_lidar}(a) and (b), resp.) show similar performances (although a slightly improvement is obtained with AR-MEM) with a global RMSE of about \SI{0.45}{\m\per\s} and a high correlation coefficient. At short integration time (Figs. \ref{fig:compare_radar_lidar}(c) and (d), resp.), however, the use of the AR-MEM allows for an important increase in the number of available measurements with respect to the conventional FFT method (from 167 to 522) as well as more accurate estimates, with a RMSE reduced by about \SI{20}{\%} and a higher correlation coefficient.

\begin{table}[h]
	\centering
	\caption{Comparison with Doppler lidar}
	\label{tab:compare}
	\begin{tabular}{ccc}
		\toprule
		Method & RMSE (\si{\m\per\s}) & Number of estimates \\
		\midrule
		FFT long & 0.45 & 64 \\
		AR-MEM long & 0.43 & 71 \\
		FFT short & 1.07 & 167 \\
		AR-MEM short & 0.81 & 522 \\
		\bottomrule
	\end{tabular} \\[1mm]
	\begin{minipage}{.81\linewidth}
		\textit{Note:} RMSE (in \si{\m\per\s}) between the colocated RWP radial velocity and the Doppler lidar measurements (see Figure \ref{fig:compare_radar_lidar}), together with the number of compared estimates for each method.
	\end{minipage}
\end{table}

\section{Discussion and Conclusion}

We have presented an application of the AR modeling with MEM to the wind profiling of the lower atmosphere using L-band Doppler radars. While the common operational approach relies on the computation of the Doppler spectrum using the FFT algorithm, we made use of an AR representation of the complex antenna voltage time-series. This allowed to bypass the time-frequency dilemma encountered when using the FFT. The MEM was further found to be efficient to estimate the AR coefficients in the context of short radar samples. We have first tested the proposed AR-MEM approach on the basis of noisy synthetic RWP time-series and quantified its augmented performances with respect to the classical spectral estimation in terms of robustness to noise when inverting the first- and second-order spectral moments from short samples.

The AR-MEM approach was applied to a \SI{24}{\hour}-set of data recorded with a RWP during an experiment in the vicinity of the landing runways of Paris CDG Airport. It was shown that the AR-MEM can be used to visualize the temporal variations of the Doppler spectrum at high temporal rate, whereas the FFT fails to accurately represents the atmospheric echo in such conditions. A statistical analysis of the performances of wind profiling using either method, AR-MEM and FFT, has been made for short-time samples using long-time samples as reference. The main outcome is that the use of AR-MEM  allows for a significant decrease in the RMSE of both first- and second-order spectral moments estimates. To further validate the AR-MEM wind profiling we analyzed a \SI{12}{\hour}-dataset of colocated RWP and lidar measurements in Payerne. Again, the AR-MEM velocity estimates were found in closer agreement to the lidar for short-time samples.

RWP measurements are often contaminated by the spurious and frequent returns of uncooperative flyers such as birds or planes. The robustness of the AR-MEM method to these parasitic echoes even with short time series has been assessed. This paves the way to new strategies to mitigate the unwanted presence of uncooperative flyers in wind profiling. Another useful application of the rapid wind profiling is the spectral analysis of clear-air turbulence echoes which requires a large variety of temporal scales. For both applications, dedicated measurement campaigns will be carried out in the future.

\section*{Acknowledgment}
\addcontentsline{toc}{section}{Acknowledgment}

The authors would like to thank Dr Maxime Hervo and the people from the Aerological station of Payerne (Federal Office of Meteorology and Climatology), Switzerland, for performing and providing RWP and lidar measurements. The authors are also indebted to Dr Philipp Currier, retired from Degreane Horizon, for his thorough advices on radar signal processing.

\bibliographystyle{abbrv}

\end{document}